\pdfoutput=1

\documentclass[pre,twocolumn,showpacs,aps,floats,floatfix]{revtex4-1}

\usepackage{natbib}
\usepackage[pdftex]{graphicx}
\usepackage{amssymb,amsmath}
\usepackage{amsfonts}
\usepackage{mathrsfs}
\usepackage[english]{babel}
\usepackage{epsfig}
\usepackage{epstopdf}
\usepackage{babelbib}
\usepackage{hyperref}
\usepackage{color}
\usepackage{tcolorbox}
\usepackage{tabularx}
\usepackage{array}
\usepackage{colortbl}
\tcbuselibrary{skins}
\usepackage{comment}
\usepackage{array}
\usepackage{bm}
\usepackage{mathtools}

\makeatletter
\newcommand*{\textoverline}[1]{$\overline{\hbox{#1}}\m@th$}
\makeatother

\begin{document}

\title{Electrostatics-driven inflation of elastic icosahedral shells as a model for swelling of viruses}
\author{An\v{z}e Lo\v{s}dorfer Bo\v{z}i\v{c}}
\affiliation{Department of Theoretical Physics, Jo\v{z}ef Stefan Institute, 1000 Ljubljana, Slovenia}
\email{anze.bozic@ijs.si}
\author{Antonio \v{S}iber}
\affiliation{Institute of physics, 10000 Zagreb, Croatia}
\email{asiber@ifs.hr}

\begin{abstract}
We develop a clear theoretical description of radial swelling in virus-like particles which delineates the importance of electrostatic contributions to swelling in absence of any conformational changes. The model couples the elastic parameters of the capsid -- represented as a continuous elastic shell -- to the electrostatic pressure acting on it. We show that different modifications of the electrostatic interactions brought about by, for instance, changes in $pH$ or solution ionic strength, are often sufficient to achieve the experimentally-observed swelling (about 10\% of the capsid radius). Additionally, we derive analytical expressions for the electrostatics-driven radial swelling of virus-like particles, which enable one to quickly estimate the magnitudes of physical quantities involved.
\end{abstract}

\date{\today}

\maketitle

\section*{Introduction}

Molecular interactions in viruses regulate their stability with respect to chemical and physical influences, such as changes in $pH$, ionic strength, and temperature. The integrity of viruses in a changing environment is important for their successful propagation from cell to cell and for their survival in the inactive, compact state in-between two hosts. At the same time, the molecular interactions involved in viral stability and integrity need to be sufficiently ``soft'' and weak in order to enable the dynamics of the viral life-cycle, which in most cases includes the disassembly of viruses, i.e., the disintegration of their protein shell (capsid) and the release of their genome (either a DNA or RNA molecule in single- or double-stranded form). These interactions are encoded primarily in the physico-chemical properties of the capsid proteins, in the nature of the packaged genome, and in some cases in the properties of proteins which serve to condense the genome and pack it more efficiently~\cite{Ceres2002,Kegel2004,Perez2015,Perlmutter2015,Roos2010}.

There is not much more to an assembled virus other than the interactions which keep it together -- once outside the host cell, the virus can be thought of as a macromolecular complex held together by electrostatic- and entropy-derived forces. And yet, the roles of these different interactions have yet to be fully elucidated. There have been attempts to study the viruses on a molecular level~\cite{Huber2017,Miao2010,Andoh2014,Freddolino2006,Zink2009,Tarasova2017}, which is a daunting task as viruses contain a huge number of atoms. For instance, even very small viruses, such as satellite tobacco mosaic virus (STMV) or southern bean mosaic virus (SBMV), contain more than $10^6$ atoms~\cite{Freddolino2006,Zink2009}. Furthermore, capsid integrity depends on the molecules which surround it, such as water and dissolved ions, which need to be taken into account in molecular simulations~\cite{Miao2010,Tarasova2017}. All this requires the knowledge of a lot of parameters which determine the different atomic interactions involved.

Other attempts at understanding the contributions of interactions involved in capsid assembly and stability use either various coarse-graining methods~\cite{Tama2002,Zandi2005,Perlmutter2015,Grime2016,Hagan2016} or simplified continuum models~\cite{Siber2012,Siber2009,Zlotnick1999,Bruinsma2003,Zandi2006}, and are based on some generalized form of interactions. This is possible because, even though the nature of capsid proteins influences the mechanics of the capsids, there exist more generic aspects of the physics of the capsid shells as well. These overarch both the chemical specificity of their protein constituents and the myriad of molecular interactions involved~\cite{Cheng2015}. The generic aspects of capsid mechanics should be prominent at the spatial scales typical for viruses ($\sim10$-$500$ nm), and it is important to clearly separate them from protein- and molecule-specific interactions. In this way, we can elucidate the background physical principles which do not depend on the molecular details, but only on a small number of parameters characterizing the elastic response of a virus shell to changes in the environment. This should be of help in identifying the wider space of physical possibilities and potentialities available in the course of the viral evolution.

What is more, coarse-grained approaches may at present be the most reliable way to extract the relevant energy scales and forces involved in viral life-cycles. Such approaches have led to important results regarding generic aspects of both electrostatic interactions in viruses~\cite{Siber2012,vdSchoot2005,Bruinsma2016} as well as their elasticity~\cite{Lidmar2003,Nguyen2005,Zandi2005,Siber2006,Siber2009,Siber2009a,Perotti2015}, and have enabled classifications of viruses according to their electrostatic ~\cite{ALB2012,ALB2018a} and elastic ~\cite{Mannige2010,May2012,ALB2013b} nature. The present study is a step further in this direction, and couples different aspects of physics of viruses in order to explain the radial expansion -- swelling -- of viruses within a simple generic framework, connecting together the relevant elastic and electrostatic energy scales.

Swelling is a quite common phenomenon in viruses~\cite{Witz2001,Belnap1996,Ausar2006}, often observed also as a side effect of the procedures applied to study the stability and (dis-)assembly of viruses~\cite{Burrage2000}. It has been studied in detail particularly in the case of cowpea chlorotic mottle virus (CCMV)~\cite{Bancroft1968,Jacrot1975,Speir1995,Konecny2006,Wilts2015}, as well as other plant viruses, such as brome mosaic virus (BMV)~\cite{Incardona1964}, STMV~\cite{Kuznetsov2001} and SBMV~\cite{Hsu1976,Rayment1979,Kruse1982}. Swelling is often triggered by the changes in the environment which modify the electrostatic interactions in the system, known to be of key importance for the fixation of the capsid shape and structure~\cite{Lavelle2009}. Swelling can thus arise due to changes in the $pH$ or the ionic strength of the solution, release of bound ions (such as Ca$^{2+}$), or different modifications of charge on the capsids.

The main aim of our work will be to elucidate the more universal physical principles that can drive capsid swelling when the latter is sufficiently small and no conformational changes occur. Depending on the stiffness of the capsid and on the magnitude of perturbation from the equilibrium state, such changes may be observable or not, so that the effect could be more prominent in some virus species than in others~\cite{ALB2013b}. In this way, we will provide a complementary view of radial swelling in capsids, and we shall explain and quantify the effects within a simple theoretical framework.

\section*{Swelling as expansion of an elastic icosahedral shell under pressure}

\begin{figure}[!tbp]
\centering{\includegraphics[width=0.7\columnwidth]{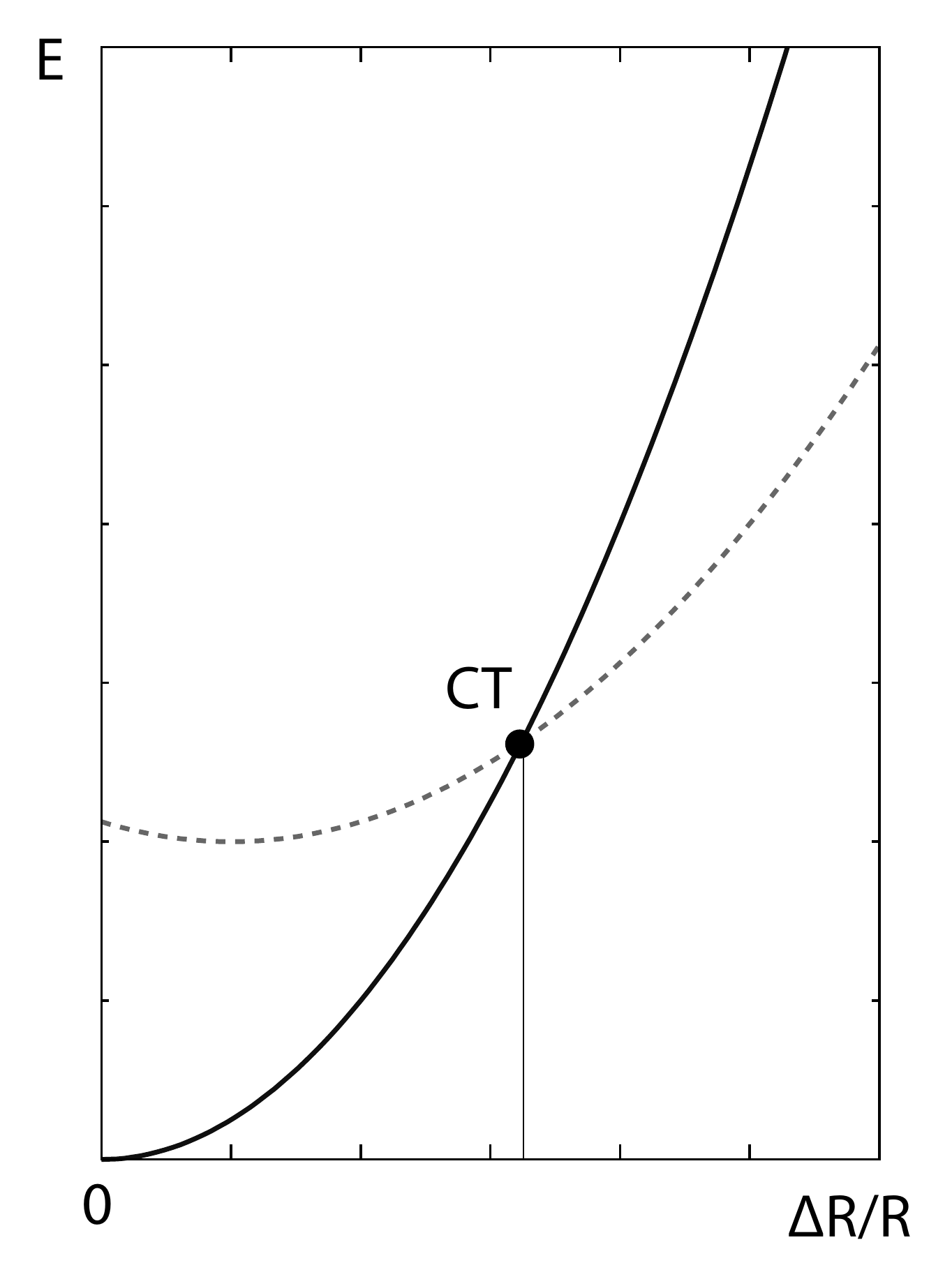}}
\caption{Effective potential energy ($E$; $y$-axis) curves in the process of capsid swelling. In cases where no conformational transition takes place, the energy increases continuously with the (relative) increase of capsid radius, $\Delta R/R$ ($x$-axis). This is depicted by the full line. In cases where protein conformational transitions occur, the energy switches from the dependence illustrated by the full line to the one illustrated by dashed line at the point of the transition (CT). The conformational change illustrated in the figure induces a softening of the capsid. In both regimes, before and after the transition, Hookean elasticity is assumed to describe the swelling of the protein shell, as can be judged from the indicated parabolic dependence of energy on $\Delta R/R$.}
\label{fig:Figure1}
\end{figure}

As a capsid can be viewed, at least approximately, as an elastic shell~\cite{Lidmar2003}, it should elastically deform in response to (small) forces acting on it. A sufficiently small swelling could thus be viewed as an elastic response of the capsid to the extra forces acting on it, a description which we will use in our work. This excludes situations where swelling involves a significant conformational change of the proteins. Such situations cannot be explained as deviations from the (elastic) equilibrium state, but rather as transitions involving at least two effective potential energy curves (Fig.~\ref{fig:Figure1}). In general, one could thus imagine that conformational changes either soften or harden the capsids, effectively modifying the capsid elastic constants and making them easier or harder to stretch upon further application of pressure. (For instance, during the swelling of CCMV, a softening of the shell can be observed~\cite{Wilts2015}.)

Conformational changes and essential modifications of the network of protein interactions during swelling transition strongly depend on the precise nature of the capsid proteins. Consequently, in order to provide a generic framework, we will neglect any conformational changes that occur during swelling. Capsids before and after a conformational change could however still be described within our framework if the change in the elastic parameters is known, as this simply causes a switch from one effective energy curve to another, as depicted in Fig.~\ref{fig:Figure1}. In a similar fashion one could include the dependence of capsid elastic constants on external, electrostatic parameters~\cite{Shojaei2016}.

Whatever the source of swelling may be, it can be represented in the lowest order as an effective internal pressure $p$. This neglects the forces which may occur tangential to the shell~\cite{Zandi2005} and which can induce tangential displacements of the proteins and their parts -- these would not be registered in the change of mean radius of the capsid, i.e., swelling. In this approximation, we view the capsid as a continuous elastic infinitely thin shell with two elastic moduli -- two-dimensional Young's modulus $Y$ and bending rigidity $\kappa$ (with Poisson ratio $\nu=1/3$, see below)~\cite{Lidmar2003,Siber2006,Siber2009,ALB2013b} -- and the swelling as a radial inflation of this elastic shell. Our model nonetheless retains the essential geometric features of virus capsids, as it possesses icosahedral symmetry and the pentagonal coordination of the points on the icosahedron vertices in otherwise hexagonally coordinated lattice of points~\cite{Lidmar2003}.

Such an elastic shell, in absence of external forcing, can be spherical to a greater or lesser degree. It has been shown that the degree of its asphericity depends on the dimensionless quantity called the F\"{o}ppl--von K\'arm\'an (FvK) number, $\gamma \equiv Y\overline{R}^2 / \kappa$, where $\overline{R}$ is the mean radius of the shell in the absence of pressure~\cite{Lidmar2003}. For $\gamma\lesssim200$, the shells are quite spherical. The icosahedron vertices pop out (buckle outwards) when $200 <\gamma < 10000$ and the shapes in this interval look like rounded icosahedra with approximately conical regions surrounding the vertices~\cite{Siber2007a}. For even larger $\gamma$, the edges of the rounded icosahedra sharpen and the shapes asymptotically approach the icosahedron~\cite{Siber2006a}.

These results can be interpreted in the framework of a continuum theory, although microscopic models of the shell elasticity can also be constructed, such as the one used in Ref.~\cite{Lidmar2003}:
\begin{equation}
E = \frac{\epsilon}{2} \sum_{i,j} \left( \left | {\bf r}_i - {\bf r}_j \right | - a \right)^2 + \frac{\tilde{\kappa}}{2} \sum_{I,J} \left( {\bf n}_I - {\bf n}_J \right)^2,
\label{eq:micro_model}
\end{equation}
where $E$ is the elastic energy of the polyhedral shell. The indices $i$ and $j$ run over the vertices of the shell (these may represent, e.g., capsomeres -- clusters of five or six proteins~\cite{Siber2006}), positioned at ${\bf r}_i$, and the indices $I$ and $J$ run over its triangular faces (plaquettes), whose normal unit vectors are denoted by ${\bf n}_I$. The equilibrium edge length is denoted by $a$, and the stretching and bending energy constants by $\epsilon$ and $\tilde{\kappa}$, respectively. The microscopic model reproduces the continuum results and elastic moduli, $Y=2 \epsilon / \sqrt{3}$ and $\kappa = \sqrt{3}\tilde{\kappa}/2$, once the number of triangular plaquettes in the shell surface becomes large enough~\cite{Lidmar2003,Siber2006}. The shapes of virus capsids have been compared to the results of the continuum theory~\cite{Lidmar2003,ALB2013b} and it was found that viruses have typically $\gamma < 10^4$~\cite{ALB2013b}. A lot of viruses are very spherical, with $\gamma<200$, but viruses with $200 <\gamma < 10^4$ are also quite numerous, and these are distinctively aspherical with visible icosahedral geometry. The model in Eq.~\ref{eq:micro_model} has also been investigated with a pressure term $pV$, where $V$ is the shell volume, added to it~\cite{Siber2006,Siber2009}, and this is also the model we adopt in this study.

The only way to produce swelling from our Hamiltonian (Eq.~\ref{eq:micro_model}) without any pressure is to increase the equilibrium length $a$ and thus obtain a less dense structure. While it is possible that the elastic parameters -- the bending and stretching constants -- change due to external parameters, these changes would not produce swelling, but only a change in shape asphericity (due to a change in $\gamma$). Thus, we perform a first-order, decoupled approximation which keeps the elastic parameters fixed during swelling and includes only pressure as its driving force. Note also that our model in Eq.~\ref{eq:micro_model} assumes a zero spontaneous curvature of the shell. However, an increase in equilibrium length $a$, which we do not consider here, could be interpreted as a decrease in spontaneous curvature. Such an approach has been elaborated in Ref.~\cite{Shojaei2016}.

When the shell is inflated and its radius increased, its elastic energy grows due to the stretching of the shell material. For perfectly spherical thin membranes, the radial force resisting the internal pressure can be obtained from the normal biaxial stresses in the material
\begin{equation}
\frac{\Delta R}{\overline{R}} = \frac{p}{2 Y}\,\overline{R}\,(1 - \nu) = 0.289\, \tilde{p},
\label{eq:estimate_baloon}
\end{equation}
where $\Delta R$ is the increase of the shell radius under pressure and $\tilde{p} = p \overline{R} / \epsilon$. In deriving Eq.~\ref{eq:estimate_baloon}, we have used $\nu = 1/3$ and assumed that $\Delta R / \overline{R} \ll 1 $. While viruses are not perfectly spherical and their mean asphericity depends on $\gamma$, the estimate in Eq.~\ref{eq:estimate_baloon} will nonetheless prove to be useful even for quite aspherical shapes, as we shall show later on. The relative increase of the shell radius is, according to Eq.~\ref{eq:estimate_baloon}, directly proportional to effective pressure, but the scale of proportionality depends on the stretching elasticity of the shell ($Y$). The analytic expression of Eq.~\ref{eq:estimate_baloon} can also be derived for large swelling, i.e., when $\Delta R / R$ is not small -- assuming that Hookean model describes the large protein displacements (see Fig.~\ref{fig:Figure1}) and that the conformational transition does not take place. This is, however, not really required, neither by the relative simplicity of our model nor by experiments, which measure small swellings~\cite{Witz2001,Speir1995,Incardona1964,Kuznetsov2001}.

\subsection*{Elastic moduli of viruses}

In order to relate the swelling of viruses to changes in their surroundings, one needs to know the Young's moduli of their capsids. In our previous study, we have found that the ratio of elastic constants $Y/\kappa$ of different viral capsids varies over four orders of magnitude, from $10^{-2}$ to $10^2$, with most viruses falling into the range $Y/\kappa \sim 0.1$-$2$ nm$^{-2}$~\cite{ALB2013b}, consistent with previous propositions ($Y/\kappa \sim 1$ nm$^{-2}$~\cite{Lidmar2003}). To extract the Young's modulus from this analysis we need an estimate of the bending rigidity. The value obtained from the analysis of aberrant assembly of empty hepatitis B capsids puts it in the range of tens of $k_B T$~\cite{Siber2009}, consistent with the value for bending rigidity used in Ref.~\cite{Nguyen2005}. This gives $Y \sim 1$-$20$ $k_B T/\mathrm{nm}^{-2}$. The bulk Young's modulus of the capsid material, obtained by dividing $Y$ with the mean capsid thickness, would be thus in the range $\sim0.5$-$10$ MPa for most viruses, two to three orders of magnitude smaller than the value obtained for bacteriophage $\phi29$~\cite{Ivanovska2004} from the analysis of AFM pressing experiments. On the other hand, our estimate is quite close to the one obtained in Ref.~\cite{Zandi2005} ($5$ MPa) in a theoretical discrete elastic study of CCMV capsids assuming effectively only a capsomere-capsomere interaction of $15$ $k_B T$.

\section*{FvK number and fixation of pressure scale}

\begin{figure*}[!tbp]
\centering{\includegraphics[width=0.6\textwidth]{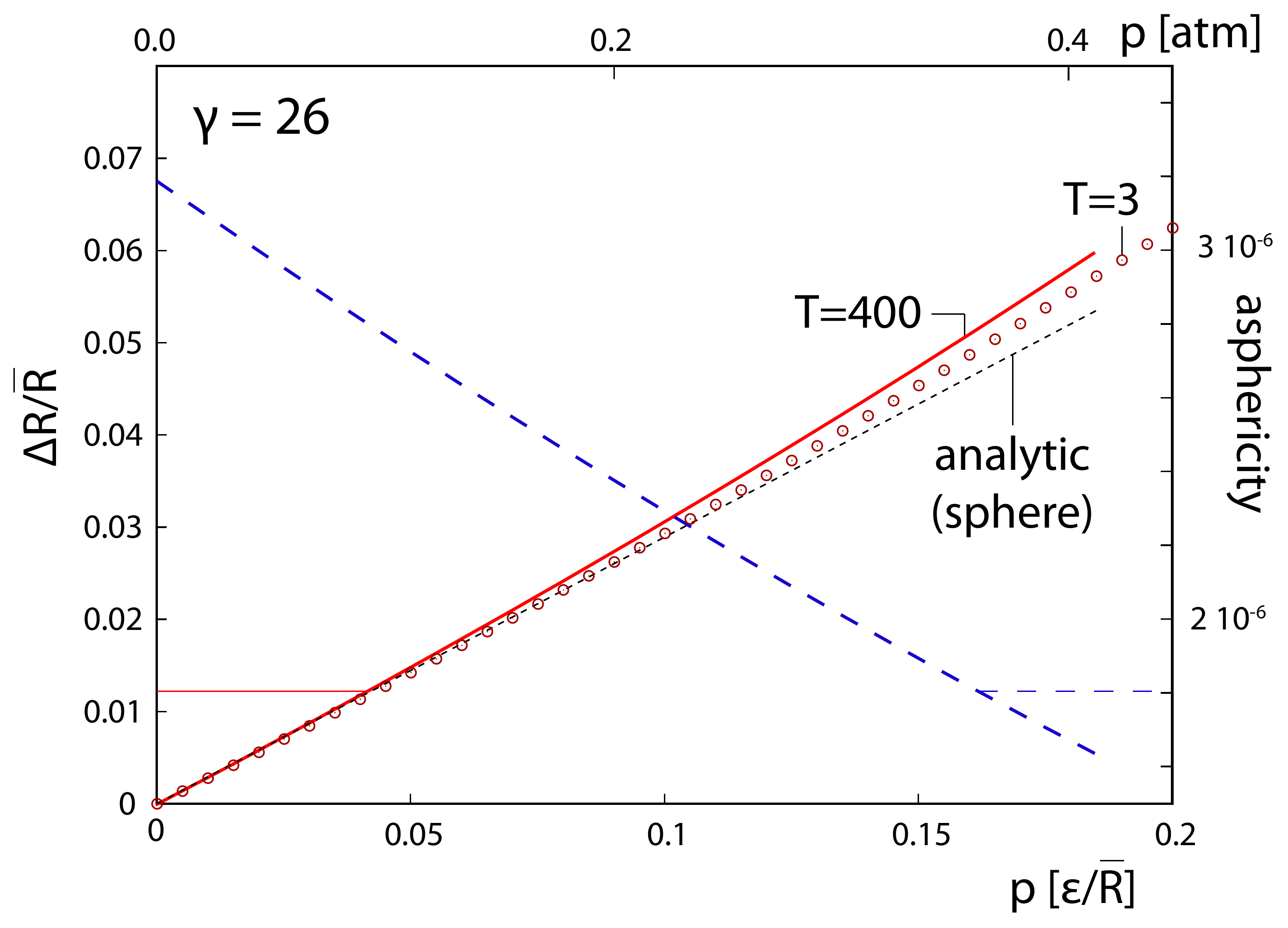}}
\caption{Mean shell radius (full line; left $y$-axis) and asphericity (long-dashed line; right $y$-axis) as a function of the internal pressure acting on a shell obtained from the numerical minimization of Eq.~\ref{eq:micro_model} with pressure~\cite{Siber2006} ($T=400$). The scaled units of pressure ($\epsilon / \overline{R}$) are shown on the bottom $x$-axis. The short-dashed line is the theoretical prediction of Eq.~\ref{eq:estimate_baloon}. The pressures recalculated in standard atmosphere (atm) units are shown on the top $x$-axis. These were obtained by taking $Y = 1$ $k_B T/\mathrm{nm}^2$, $\kappa = 10$ $k_B T$, and $\overline{R} = 16$ nm, yielding $\gamma = 26$, which is the FvK number at which the calculations were performed. The results of a numerical minimization of energy of a shell with $T=3$ and same $\gamma$ are indicated by circles.}
\label{fig:Figure2}
\end{figure*}

As mentioned in the previous Section, the outcome of elastic calculations in the continuum limit depends only on the combined quantity of the FvK number $\gamma$. However, at a fixed ratio of elastic constants $Y/\kappa$, a given FvK number also implies a fixed shell radius. Figure~\ref{fig:Figure2} shows the characteristics of shell shape -- its mean radius and asphericity~\cite{Siber2006} -- under internal pressure, for $\gamma = 26$ (in the non-pressurized state). The internal pressure is scaled with $\epsilon$ rather than with $Y$ so to enable more direct comparisons with the results from Ref.~\cite{Siber2006}. Taking $Y = 1$ $k_B T/\mathrm{nm}^2$ and $\kappa = 10$ $k_B T$, consistent with the interval of values found for viruses, gives then $\overline{R} = 16$ nm for this choice of $\gamma$. As $Y$ (or, equivalently, $\epsilon$) fixes the scale of pressure, this enables a transition from scaled units of pressure shown in the bottom $x$-axis of Fig.~\ref{fig:Figure2} to SI units shown in the top $x$ axis. A pressure of about half an atmosphere in this case induces an increase in mean radius of about 7\% -- comparable to expansions observed experimentally in STMV (8\%)~\cite{Kuznetsov2001}, CCMV (10\%)~\cite{Speir1995}, BMV (12\%)~\cite{Incardona1964}, and TBSV (14\%)~\cite{Witz2001}.

The calculations performed here should be viewed as a continuum limit of the model elaborated in Ref.~\cite{Lidmar2003}, as they are performed for a triangulation number ${\cal T}=400$ ($h=20$, $k=0$), which is already a very good representation of a continuum situation, as demonstrated in Ref.~\cite{Siber2006}. However, the continuum limit is almost reached already at very small ${\cal T}$-numbers, as the calculations for ${\cal T}=3$ in Fig.~\ref{fig:Figure2} clearly show. This further strengthens the applicability of our results and the simple estimate of Eq.~\ref{eq:estimate_baloon} to real viruses, i.e., small ${\cal T}$-numbers. To obtain the same $\gamma$ with a shell of a smaller ${\cal T}$-number requires either a change of equilibrium edge length $a$ so as to obtain a similar mean radius $\overline{R}$, or an increase (decrease) of $Y$ ($\kappa$). We have checked that the different choices produce virtually indistinguishable results, as long as they produce the same $\gamma$, even in the case where a small shell, far from the continuum limit, is studied~\cite{Siber2006}.

For $\gamma$ as small as the one used in Fig.~\ref{fig:Figure2} ($\gamma = 26$), the shell is nearly a perfect sphere. Although the drop in asphericity is obtained in the calculations, the overall asphericity, as defined in Ref.~\cite{Lidmar2003}, remains rather small ($\sim$ 10$^{-6}$), and the shell is practically a perfect sphere both in the native and in the swollen form. The prediction of Eq.~\ref{eq:estimate_baloon} agrees with the numerical results quite well, especially when $\Delta R / \overline{R} < 0.02$. The calculations thus validate the simple relation between the magnitude of swelling and the effective pressure, but still require confirmation in the case of more aspherical shells.

\begin{figure*}[!tbp]
\centering{\includegraphics[width=0.6\textwidth]{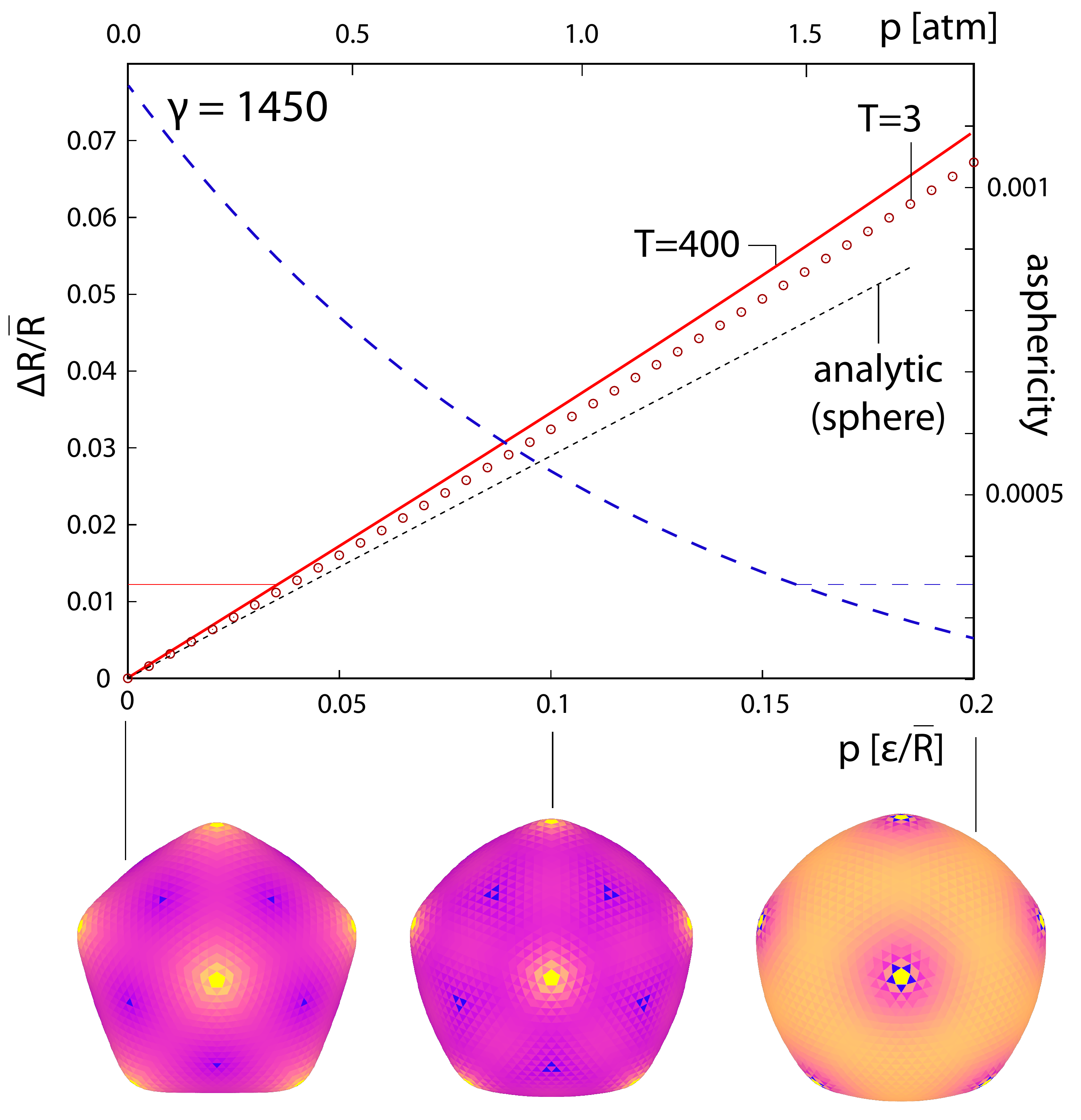}}
\caption{Mean shell radius (full line; left $y$-axis) and asphericity (long-dashed line; right $y$-axis) as a function of the internal pressure acting on a shell obtained from the numerical minimization of Eq.~\ref{eq:micro_model} with pressure~\cite{Siber2006} ($T=400$). The scaled units of pressure ($\epsilon / \overline{R}$) are shown on the bottom $x$-axis. The pressures recalculated in standard atmosphere (atm) units are shown on the top $x$-axis. These were obtained by taking $Y = 10$ $k_B T/\mathrm{nm}^2$, $\kappa = 10$ $k_B T$ and $\overline{R} = 38$ nm, yielding $\gamma = 1450$, which is the FvK number where the calculations were performed. The results of a numerical minimization of energy of a shell with ${\cal T}=3$ and same $\gamma$ are indicated by circles. The three shapes shown below the bottom $x$-axis were calculated for $p=0$, $0.1$, $0.2$ $\epsilon / \overline{R}$ ($T=400$), as indicated in the figure. The brightly colored (yellow in the online version) shell regions store relatively larger elastic energy than the darkly colored (violet in the online version) regions.}
\label{fig:Figure3}
\end{figure*}

The so-called (outward) ``buckling transition'' of a thin elastic shell takes place when $\gamma \sim 10^3$, as the pentagonal disclinations buckle out and the shell becomes significantly aspherical. It is of interest to see how the internal pressure changes the buckled shell geometry, and Fig.~\ref{fig:Figure3} shows the calculation for $\gamma=1450$. The scale of pressure (top $x$-axis) in this case is chosen differently than in Fig.~\ref{fig:Figure2} to correspond to the mid-range of the interval of elastic ratios found in Ref.~\cite{ALB2013b}. The values chosen are $Y = 10$ $k_B T/\mathrm{nm}^2$, $\kappa = 10$ $k_B T$ ($Y / \kappa = 1$ nm$^{-2}$) and $\overline{R}=38$ nm, and they yield $\gamma = 1450$, the FvK number for which the numerical calculations were performed. In this case, a notable decrease in asphericity occurs as the internal pressure increases, from $\sim 10^{-3}$ at zero pressure to about $\sim 3\times 10^{-4}$ at $p=2$ atm. The rounding of the ``inflated'' capsid as the pressure increases can also be observed in the three shapes shown in Fig.~\ref{fig:Figure3}. The theoretical prediction of Eq.~\ref{eq:estimate_baloon} remains quite reliable, even though the non-pressurized shape of the shell is not a sphere anymore. The slope of the analytical $\Delta R / \overline{R}$-$\tilde{p}$ dependence ($0.289$) does not correspond to numerical results when $\Delta R / \overline{R} \rightarrow 0$, but it can still be used to quickly and reliably estimate the magnitudes of physical quantities involved, as it underestimates the exact numerical results by only about 20\%.

\subsection*{Basic electrostatic pressure estimation}

Capsid swelling and shape transitions are often observed when the electrostatic interactions acting on the capsid are modified -- for instance, when $pH$ or salt concentration of the surrounding solution are changed. Importantly, modifications of electrostatic interactions can result in significant changes of the electrostatic pressure even when they do not incur conformational changes in the capsid proteins~\cite{Siber2012}. For a homogeneously charged thin spherical shell of radius $R$, the electrostatic pressure acting on it can be obtained by deriving its free energy with respect to the volume. In the Couloumb limit -- in the absence of salt ions -- the resulting pressure is $p_C=3\sigma^2/2\varepsilon\varepsilon_0$~\cite{Siber2007}, and is clearly independent of the capsid radius. Here, $\sigma$ is the surface charge density of the capsid and $\varepsilon=80$ the dielectric constant of water. When we add a monovalent $1:1$ salt of concentration $c_0$, we obtain for the pressure in the Debye-H\"uckel (DH) regime~\cite{ALB2013a}
\begin{equation}
\label{eq:dh}
p_{DH}=p_C\times\frac23\left[\frac{1}{\kappa_{DH} R\,(1+\coth\kappa_{DH} R)}+\frac{e^{-2\kappa_{DH} R}}{2}\right],
\end{equation}
where $\kappa_{DH}^{-1}=\sqrt{\varepsilon\varepsilon_0/2\beta e_0^2c_0}$ is the inverse DH screening length. In the limit of vanishing salt, we indeed obtain $\lim_{\kappa_{DH} R\to0}p_{DH}=p_C$. On the other hand, when the screening becomes large, $\kappa_{DH} R\to\infty$, the electrostatic pressure vanishes. From Eq.~\ref{eq:dh} one can see that the two important factors in determining the electrostatic pressure on the capsid are the screening (in the form of $\kappa_{DH} R$) and the surface charge density $\sigma$, which is in case of viruses typically in the range of $\vert\sigma\vert\lesssim0.5$ $e_0/$nm$^2$ at neutral $pH$~\cite{ALB2012,ALB2018a}.

\begin{figure}[!tbp]
\centering{\includegraphics[width=\columnwidth]{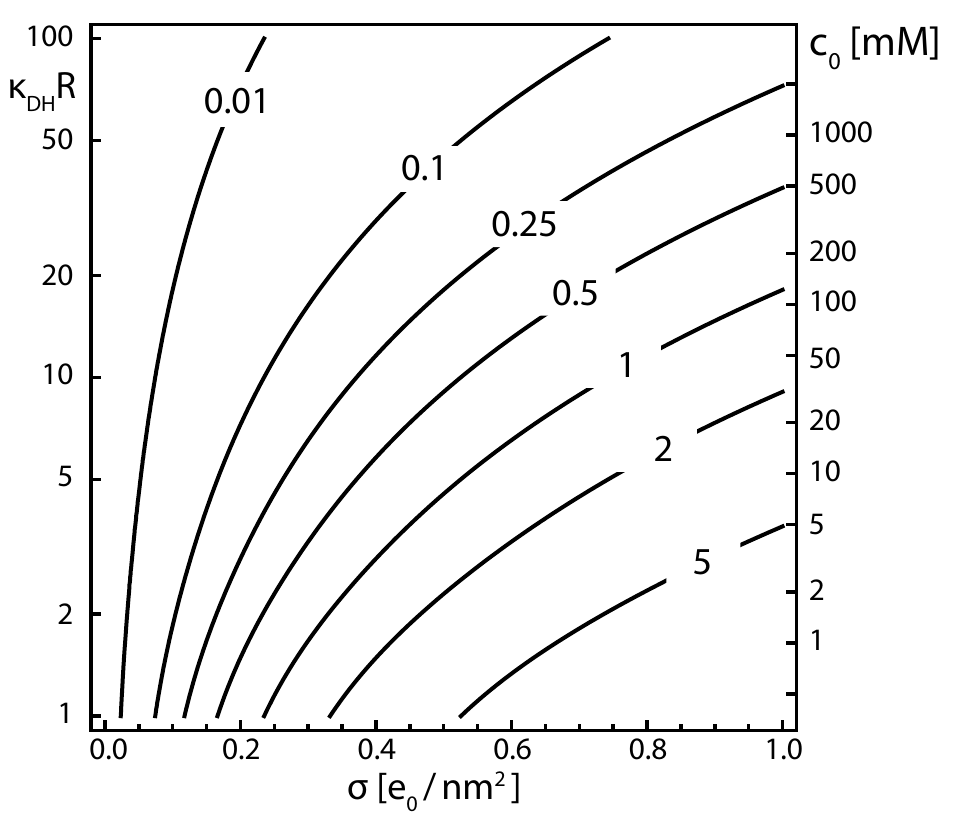}}
\caption{Isolines of DH pressure (Eq.~\ref{eq:dh}; in atm units) as a function of scaled screening length, $\kappa_{DH}  R$ (left $y$-axis) and surface charge density $\sigma$ ($x$-axis). The right $y$-axis shows the monovalent salt concentration $c_0$ in the particular case when the shell has a radius of $R=16$ nm.}
\label{fig:Figure4}
\end{figure}

Figure~\ref{fig:Figure4} shows how the DH electrostatic pressure depends on the screening length scaled by shell radius, $\kappa_{DH} R$ (left $y$-axis), and surface charge density ($x$-axis). As already shown, the pressures required to achieve notable swelling depend on the capsid elastic parameters and its radius, through a combination yielding a FvK number $\gamma$ (Figs.~\ref{fig:Figure2} and~\ref{fig:Figure3}). The shell radius is an important factor also in determining the electrostatic pressure, which depends quite strongly on the scaled screening length of the system -- left $y$-axis in Fig.~\ref{fig:Figure4}. In general, pressures that can drive radial swelling can be obtained at larger salt concentrations when the shell radii are small, and shells with large radii require very low salt concentrations in order to achieve significant pressures. The largest electrostatic pressure is reached in the Coulomb limit, having relevance for osmotic shock experiments when salt is added or depleted from a solution. Another way to achieve large changes in electrostatic pressure, even at large monovalent salt concentrations, is by adding polyvalent ions to the solution~\cite{Javidpour2013a}; in such cases, however, analytical estimates for the pressure are difficult to derive. What Fig.~\ref{fig:Figure4} clearly shows is that weakening of the electrostatic screening can lead to increases in electrostatic pressure large enough to drive radial swelling of capsids~\cite{Burrage2000}.

Another important factor determining the electrostatic pressure acting on the capsid is its surface charge density. The most common way of changing the charge on the capsid is by changing the $pH$ value of the solution, thus regulating the charge on the constituent amino acids of the capsid proteins~\cite{Lavelle2009,Nap2014,ALB2018a}. A study by Nap et al.~\cite{Nap2014} examined the $pH$ dependence of the capsid charge of several bacteriophages, all of which exhibited similar properties with the charge on the capsid showing several large changes when the $pH$ shifted from acidic to basic (see Fig.~7 in Ref.~\cite{Nap2014}). The changes in total charge in the studied bacteriophages range around $\Delta\sigma=0.1$-$0.2$ $e_0/$nm$^2$, and can be even larger in viruses whose total charge is larger at neutral $pH$. Consequently, $pH$-induced changes in surface charge density can be large enough to induce electrostatic swelling (cf.\ Figs.~\ref{fig:Figure4} and~\ref{fig:Figure2}). Of course, we assume here that no conformational changes occur in capsid proteins when the $pH$ is changed, which is often not the case. However, our results indicate that the pressure-driven radial swelling can reach similar magnitudes to swelling in the presence of capsid conformational changes.

There are also other ways to modify the surface charge density of capsids. One way involves mutation of the amino acids in the capsid protein, where mutants with as many as $4$ added or removed positive charges per capsid protein have been made~\cite{Ni2012,Hema2010}. For instance, the addition of two fully ionized residues per capsid protein on a capsid of $\overline{R}=16$ nm and $\mathcal{T}=3$ leads to a change of $\Delta\sigma=180\,\mathcal{T}e_0/4\pi\overline{R}^2\sim0.17$ $e_0/$nm$^2$. Yet another mechanism for the modification of the capsid surface charge density is the adsorption of divalent ions and their removal -- chelation. Chelation often leads to capsid swelling, albeit one which is usually related to protein conformational changes~\cite{Aramayo2005,Sherman2006,Llauro2015,Witz2001}. Divalent ions modulate the mechanics of viral capsids, regulated by the large electrostatic forces imparted by the ions. For instance, it was estimated that the virion of red clover necrotic mosaic virus -- a $\mathcal{T}=3$ virus with $\overline{R}\sim15$ nm -- has approximately $390\pm30$ Ca$^{2+}$ ions bound to the capsid~\cite{Sherman2006}. This amounts to about 2 bound divalent ions per capsid protein, imparting an additional surface charge density of $\Delta\sigma\sim0.25$ $e_0/$nm$^2$. While the adsorption and localization of Ca$^{2+}$ ions is an ion-specific effect, often resulting in strong conformational changes, it also has a non-specific background resulting simply from the large additional charge brought on by the ions. This non-specific effect of the adsorption of Ca$^{2+}$ ions certainly contributes to swelling and can be evaluated within our model.

Decrease of the salt concentration enhances (de-screens) the electrostatic repulsion and increases the electrostatic pressure acting on the capsid shell. This should be a continuous effect, unlike the changes in $pH$ which result in an essentially discrete jump in capsid surface charge density, $\Delta\sigma$~\cite{Nap2014}. This effect has been noted in the literature in the case of polyovirus, whose radius increases from $28$ nm in $100$ mM Tris to $31$ nm in $10$ mM Tris, and further to $34$ nm in $1$ mM Tris. Such a large radial swelling is not inconsistent with our data. If we assume that the virus has an effective charge density of about $0.5$ $e_0/$nm$^2$, the decrease in Tris concentration from $100$ mM to $1$ mM results in an increase of electrostatic pressure from about $0.2$ atm to $1.5$ atm (Fig.~\ref{fig:Figure4}), which may then lead to strong swelling depending on the stretching elasticity (Figs.~\ref{fig:Figure2}-\ref{fig:Figure4}). 

\section*{Discussion and conclusions}

We have presented a study of virus swelling in a generic elastic-electrostatic framework, which treats swelling as an inflation of an elastic icosahedral shell under pressure. The calculations presented in Figs.~\ref{fig:Figure2} and~\ref{fig:Figure3} -- although (deliberately) simple when compared to real viruses -- reach far enough to determine the characteristic pressure scale required for capsid swelling. We have found that the necessary pressures required to reach the experimentally-observed swelling of about 10\% fall into the range of $p\sim0.5$-$5$ atm. This estimate is based on elastic properties of the viruses quantified in several previous works~\cite{ALB2013b,Lidmar2003,Zandi2005,Siber2009}. We have also shown that such pressures can be realized by the changes in ionic screening and capsid surface charge density which can be induced experimentally.

The two-dimensional nature of our model may appear as an oversimplification when compared to real viruses. However, the atomically inhomogeneous capsid should be imagined as a network of points -- those representing ``soft'' regions of proteins and the points of contact between the proteins. The elastic response involved in the radial expansion of the capsid is encoded in the energetics of these contacts, while all the other protein regions can be treated as essentially fixed, as they are much more difficult to stretch. Thus, perhaps somewhat paradoxically, a 2D model may be a better approximation of a real capsid than the one which represents it as a thick shell. In such a model, the shell thinning upon expansion would appear as an important effect~\cite{Wilts2015}, modifying the effective response of the shell; this feature is not present in our approach. Note also that our model does not require specification of the soft and hard regions of the shell, and takes the same form, irrespectively of whether the regions which stretch are within the capsomeres or between them.

While our model decoupled the elastic and electrostatic contributions to swelling, in real systems a change in conditions is likely to produce both an effective pressure together with a change in the capsid elastic parameters. This could be caused either by conformational changes influencing the equilibrium length $a$ of the capsid protein network (Eq.~\ref{eq:micro_model}), thus leading for instance to a less dense structure, or by the changes in the electrostatic interactions in the system~\cite{Shojaei2016} modifying the shell elasticity and equilibrium lengths. However, the renormalization of the capsid elastic parameters brought about by these effects has not been studied in detail and is difficult to determine precisely. Our approach thus presents a sort of a decoupling approximation, i.e. the elastic properties of the shell are assumed to be unchanged in the process of swelling which is driven exclusively by the excess pressure.

One may also wonder whether the representation of a virus as a homogeneously charged shell is an oversimplification, and how the (attractive) electrostatic interactions between the patches with predominantly positive and predominantly negative charge are treated in this approach. Note, however, that the variations in positive and negative charge repeat in each capsomere of the capsid, so that the interactions affected by these variations are mostly tangential, the strongest contribution coming from the charge in a particular capsomere and those around it. These, thus, do not contribute to swelling, which is mainly governed by the overall imbalance of charge and the global predominance of either positive or negative charge. In our approximation, the total charge is simply smeared homogeneously (continuously) over the entire capsid surface. One could also perform a multipole expansion of the complete surface charge density, as done, for instance, in Refs.~\cite{Marzec1993} and~\cite{ALB2013a}. However, higher-order corrections (pertaining to higher-order multipoles) fall off more quickly with the screening parameter $\kappa_{DH}R$; what is more, due to the icosahedral symmetry of viral capsids, the first multipole correction can occur only at the multipole with the wave number $\ell=6$~\cite{ALB2013a}.

Our simplified model does not explicitly account for the presence of the viral genome, either single- or double-stranded RNA or DNA. Although it may seem that its applicability is thus reduced only to empty viruses, it can be applied to filled RNA viruses as well, presuming that the changes in the RNA distributions accompanying swelling occur on an energy scale much softer than the one pertaining to capsid proteins. In such cases, only the protein-protein interactions essentially determine the amount of swelling~\cite{Siber2012}, but one should note that the presence of charged ssRNA modifies the charge equilibrium in the system~\cite{Siber2008,Erdemci2014,Erdemci2016}. Such reasoning appears to be corroborated to some extent by AFM pressing experiments performed on CCMV~\cite{Wilts2015}, which measure similar, although discernibly different elastic responses of filled and empty viruses. Our description is particularly relevant for functional viruses which feature RNA distributed in a shell close to the capsid, held there by the basic tails of capsid proteins~\cite{Siber2012,ALB2018a}. The two ``shells'' of charge -- that of the RNA and that of the capsid -- can in the lowest order be treated as a single shell of excess charge, so that the approach we presented here can be applied quite directly. On the other hand, while it has recently been shown experimentally that the presence of dsDNA genome in viral capsids effectively modifies the detectable surface charge density of the capsid~\cite{Merche2015}, this is not the major effect DNA genome has on the virion. The capsids of dsDNA viruses and bacteriophages in particular already suffer a significant internal pressure built up by the confinement of the DNA. This pressure, which can be as high as tens of atmospheres~\cite{Siber2012}, can be additionally modified by changes in the electrostatic interactions. However, this generic mechanism that we have elaborated here would modify the overall pressure only to a relatively small extent, and has consequently less significance for dsDNA viruses.

Our results, based on a very generic mechanism which does not depend on the details of the capsid proteins, can also prove useful in the design of $pH$-responsive nanoparticles use in, e.g., drug delivery~\cite{Gao2010,Liu2014,Pillai2013}, where swelling, dissociating, or surface charge switching can be controlled by $pH$ in a manner that favors drug release at the target site over surrounding tissues. What is more, synthetic protein nanocontainers of non-viral origin often have similar material properties to viral ones~\cite{Heinze2016}, and can be computationally designed to have desired properties for drug delivery and other biomedical applications~\cite{Butterfield2017}.

\section*{Acknowledgments}

We thank M.\ Comas-Garcia and R.\ Podgornik for helpful comments on the manuscript. ALB acknowledges the financial support from the Slovenian Research Agency (research core funding No.\ (P1-0055)).

\bibliography{references}

\end{document}